\documentclass[aps,pra,twocolumn,superscriptaddress,showpacs]{revtex4}
\usepackage{amsmath}
\usepackage{epsfig}

\newcommand{\beq}{\begin{eqnarray}}
\newcommand{\eeq}{\end{eqnarray}}

\begin{document}

\title{Detecting non-Markovian plasmonic band gaps in quantum dots using
electron transport}
\author{Yueh-Nan Chen}
\email{yuehnan@mail.ncku.edu.tw}
\affiliation{Department of Physics and National Center for Theoretical Sciences, National
Cheng-Kung University, Tainan 701, Taiwan}
\author{Guang-Yin Chen}
\affiliation{Institute of Physics, National Chiao Tung University, Hsinchu 300, Taiwan}
\author{Ying-Yen Liao}
\affiliation{Department of Applied Physics, National University of Kaohsiung, Kaohsiung
811, Taiwan}
\author{Neill Lambert}
\affiliation{Advanced Science Institute, The Institute of Physical and Chemical Research
(RIKEN), Saitama 351-0198, Japan}
\author{Franco Nori}
\affiliation{Advanced Science Institute, The Institute of Physical and Chemical Research
(RIKEN), Saitama 351-0198, Japan}
\affiliation{Physics Department and Center for Theoretical Physics, The University of
Michigan, Ann Arbor, M1 48109-1040, USA}
\date{\today}

\begin{abstract}
Placing a quantum dot close to a metal nanowire leads to drastic changes in
its radiative decay behavior because of evanescent couplings to surface
plasmons. We show how two non-Markovian effects, band-edge and retardation,
could be observed in such a system. Combined with a quantum dot \textrm{p-i-n%
} junction, these effects could be readout via current-noise measurements.
We also discuss how these effects can occur in similar systems with
restricted geometries, like phononic cavities and photonic crystal
waveguides. This work links two previously separate topics: surface-plasmons
and current-noise measurements.
\end{abstract}

\maketitle

%%%%%%  VERSION 16 JUNE 2006 - as submitted
%\preprint{}



%\preprint{APR 2004-XX}

\section{Introduction and motivation}

When a photon strikes a metal surface, a surface plasmon-polariton (a
surface electromagnetic wave that is coupled to plasma oscillations) can be
excited. The concept of plasmonics$^{1}$, in analogy to photonics, has
arisen as a new and exciting field since surface plasmons reveal strong
analogies to light propagation in conventional dielectric components$^{2}$
and provide a possible miniaturization of existing photonic circuits$^{3}$.

In a related context, a complete understanding of the dynamics of quantum
systems interacting with their surroundings has become desirable,
particularly with respect to applications for quantum information science.
While the Markovian approximation is widely adopted to treat decoherence and
relaxation problems, the non-Markovian dynamics of qubit (two-level) systems
have come under increased scrutiny$^{4}$. This is because a simple Markovian
description is not adequate when the qubit is strongly coupled to its
environment. In solid state systems, an exciton in a quantum dot (QD) can be
viewed as such a two-level system. Recently single-qubit gate operations on
QD excitons have been studied experimentally$^{5}$. Furthermore, with
advances in fabrication technologies, it is now possible to embed QDs inside
a \textrm{p-i-n} structure$^{6}$, such that electrons and holes can be
injected separately from opposite sides. This allows one to examine the
exciton dynamics in a QD via electrical currents$^{7}$.

Motivated by these recent developments in plasmonics and quantum information
science, we show in this work how non-Markovian interactions between QD
excitons and nanowire surface plasmons give rise to two interesting effects:
band-edge and retardation. In a different system, the band-edge effect was
originally predicted using the isotropic band-edge model$^{8}$: the
quadratic dispersion relation, $\omega _{k}=\omega _{c}+A(k-k_{c})^{2}$,
leads to a photonic density of state $\rho (\omega )$\ at a band-edge $%
\omega _{c}$, which behaves as $1/\sqrt{\omega -\omega _{c}}$\ for $\omega
\geq \omega _{c}$. In a nano-wire, the band-edge effect stems from the
non-linear behavior of the plasmon dispersion relation, in which there are
similar quadratic local extremes at certain frequencies. The other effect we
investigate here, retardation, is the multiple time delay of emission and
absorption of plasmons between two QDs. With the incorporation of the system
inside a \textrm{p-i-n}\textit{\ }junction, we show that both effects can be
readout via current-noise measurements. The possibility of observing such
phenomena in a QD spin qubit confined in a \emph{phononic} cavity or a QD in
a \emph{photonic} crystal waveguide are also discussed.
\begin{figure}[h]
\includegraphics[width=\columnwidth]{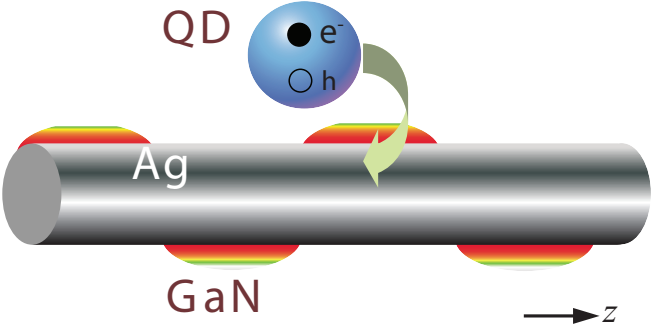}
\caption{{}(Color online) Schematic view of the system: a metallic
(e.g., silver here) nano-wire is embedded inside a GaN matrix and a (blue) QD (quantum dot) is placed on top of
it. An evanescent electromagnetic wave couples the metallic wire and
the QD. The exciton in the QD (presented by the two disks) can
recombine, spontaneously emitting photons (green arrow) that produce
surface plasmons on the wire (illustrated by the surface effect). }
\end{figure}

\section{Band-edge effect}

Consider now a semiconductor QD near a cylindrical metallic (we will
consider silver here) nanowire with radius $a$ and longitudinal axis $z$ as
shown in Fig.~1. The QD and nanowire are assumed to be separated by a
dielectric layer$^{9}$. The \emph{n-}th surface plasmon mode's components of
the electromagnetic field at the surface can be obtained by solving
Maxwell's equations in a cylindrical geometry ($\rho ~$and $\varphi $ denote
the radial and azimuthal coordinates, respectively) with appropriate
boundary conditions$^{10}$. The dispersion relations of the surface plasmons
can be obtained by numerically solving the following transcendental equation:

\begin{align}
& S(k_{z},\omega )=  \notag \\
& \left[ \frac{\mu _{I}}{K_{I}a}\frac{J_{n}^{\prime }(K_{I}a)}{J_{n}(K_{I}a)}%
-\frac{\mu _{O}}{K_{O}a}\frac{H_{n}^{(1)\prime }(K_{O}a)}{H_{n}^{(1)}(K_{O}a)%
}\right] \times   \notag \\
& \left[ \frac{(\omega /c)^{2}\varepsilon _{I}(\omega )}{\mu _{I}K_{I}a}%
\frac{J_{n}^{\prime }(K_{I}a)}{J_{n}(K_{I}a)}-\frac{(\omega
/c)^{2}\varepsilon _{O}(\omega )}{\mu _{O}K_{O}a}\frac{H_{n}^{(1)\prime
}(K_{O}a)}{H_{n}^{(1)}(K_{O}a)}\right]   \notag \\
& -n^{2}k_{z}^{2}\left[ \frac{1}{(K_{O}a)^{2}}-\frac{1}{(K_{I}a)^{2}}\right]
^{2}  \notag \\
& =0,
\end{align}%
whose solutions are the dispersion relations $\omega _{n}=\omega _{n}(k_{z})$%
. Here, \emph{I} (\emph{O}) stands for the component inside (outside) the
wire. Also, $J_{n}(K_{I}\rho )$ are $H_{n}^{(1)}(K_{O}\rho )$ are the Bessel
and Hankel functions, respectively. The dielectric function is assumed as
\begin{equation}
{\epsilon }(\omega )=\varepsilon _{\infty }\left[ 1-\frac{\omega _{p}^{2}}{%
\omega (\omega +i/\tau )}\right] ,
\end{equation}%
where $\epsilon _{\infty }=9.6$ (for Ag), $\epsilon _{\infty }=5.3$ (for
GaN), $\omega _{p}$ is the plasma frequency, and $\tau $ is the relaxation
time due to ohmic metal loss$^{11}$. The magnetic permeabilities $\mu _{I}$
and $\mu _{O}$ are unity everywhere since here we consider nonmagnetic
materials. The reason to choose a silver nanowire here is that the plasmon
energy $\hbar \omega _{p}$ of bulk silver is $3.76~eV$ with the
corresponding saturation energy $\hbar \omega _{p}/\sqrt{2}\approx 2.66eV$
in the dispersion relation. As we shall see below, variations of the
dispersion relations in energy just match the exciton bandgap of wide-band-gap
nitride semiconductor QDs.  In related work, Gallium nitride is used as a matrix interface between a silver film
and a indium gallium nitride quantum well$^{12}$.  This is primarily because the refractive index of GaN reduces the surface plasmon energy to
match that of the exciton energy. 
\begin{figure}[h!t]
\includegraphics[width=7.5cm]{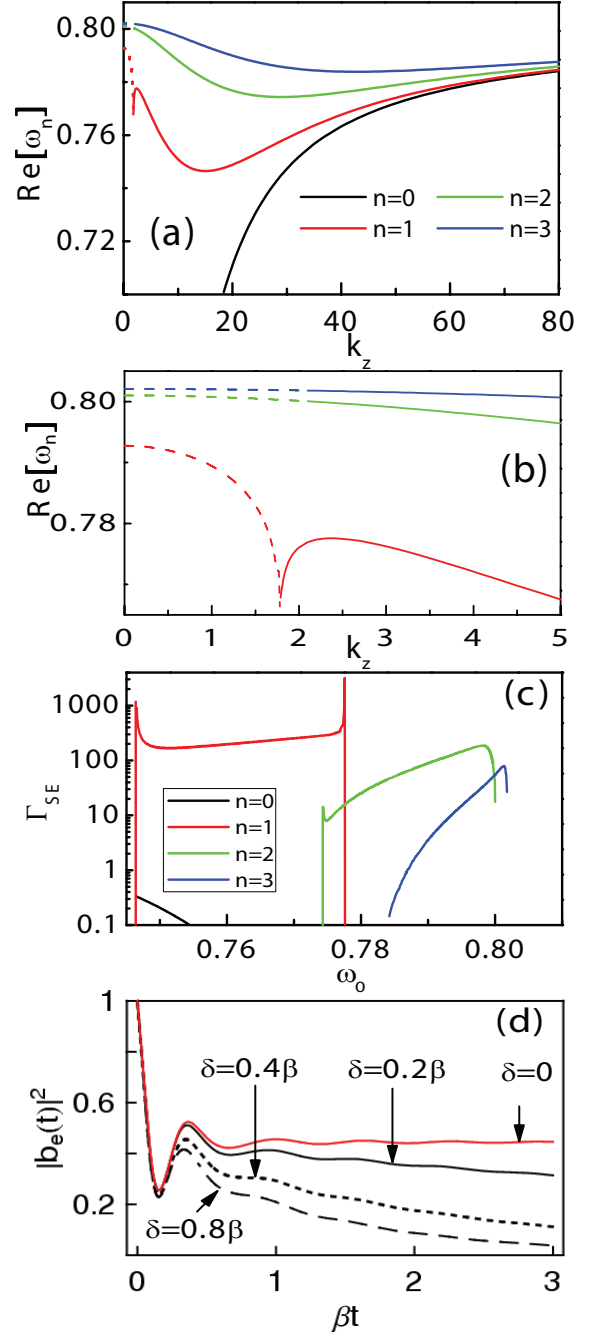}%width=8cm
\caption{(Color online) (a) Dispersion relations $Re[\protect\omega
_{n}]$ versus $K_{z}$ of surface plasmons for the first few modes
($n=0,1,2,3$).
The units for the vertical and horizontal lines are$\ \Omega =\protect\omega %
/\protect\omega _{p}$ and $K=k_{z}c/\protect\omega _{p}$. (b) The
enlarged plot of the dispersion relations of (a) in the regime of
small $k_{z}$. (c) Corresponding (Markovian) spontaneous emission
(SE) rates into surface plasmons. As seen here, the SE rates are
greatly enhanced at certain values of $\protect\omega _{0}$.\emph{\
}(d) Non-Markovian decay dynamics of QD
excitons for $\protect\delta =0.2\protect\beta $ (black line)$,$ $0.4\protect%
\beta $ (dotted line)$,$ and $0.8\protect\beta $ (dashed line). When $%
\protect\delta =0,$ the red curve represents the result of the
contribution from the $n=1$ mode.}
\end{figure}

The dispersion relations for various modes $n$ are shown in Fig. 2(a) with
effective radius $R=0.1$. The unit of the effective radius $R$ ($\equiv
\omega _{p}a/c$) is roughly equal to $53.8$ nm. The behavior of the $n=0$
mode is very similar to the two-dimensional case$^{13}$, i.e. $\Omega $
gradually saturates with increasing wave vector $k_{z}$. This is because the
fields for the $n=0$ mode are independent of the azimuthal angle $\varphi $.
However, the behavior for the $n\neq 0$ modes are quite different. The first
interesting point are the discontinuities around $\omega /c\approx k_{z}$.
Further analysis shows that the solutions of $\omega $ are ``almost real''$%
^{14}$ when $k_{z}>Re[\omega ]/c$. Thus, the first Hankel function of order
\emph{n}, $H_{n}^{(1)}(K_{\xi }\rho )$, decays exponentially. This means
that the surface plasmons in this regime are confined to the surface (%
\textit{bound modes}). For $k_{z}$ $<Re[\omega ]/c$, however, the solutions
of $\omega $ are \emph{complex}, as shown by the dashed lines in Fig. 2(b). $%
H_{n}^{(1)}(K_{\xi }\rho )$ in this case is like a traveling wave with
finite lifetime (\textit{non-bound modes}).

\subsection{Spontaneous emission rates}

Once the electromagnetic fields are determined, the spontaneous emission
(SE) rate, $\Gamma _{SE}$, of the QD excitons into bound surface plasmons
can be obtained via Fermi's golden rule. The SE rates of the first few modes
($n=0,1,2,3$) are shown in Fig. 2(c) with effective radii $R=0.1$. In
plotting the figures, the distance between the dot and the wire surface is
fixed as $d=10.76~$nm. The novel feature is that the SE rate approaches
infinity at certain values of the exciton bandgap $\omega _{0}$.
Mathematically, one might think that at these values the corresponding
slopes of the dispersion relation are zero. Physically, however, this
infinite rate is not reasonable since it is based on perturbation theory.
Therefore, one has to treat the dynamics of the exciton around these values
more carefully, i.e. the \emph{Markovian} SE rate is not enough. One has to
consider the \emph{non-Markovian} behavior around the band-edge, which means
the band abruptly appears/disappears across certain values of $\omega $.

\subsection{Non-Markovian dynamics}

To obtain the non-Markovian dynamics of the exciton, we first write down the
Hamiltonian of the system in the interaction picture (within the rotating
wave approximation),

\begin{eqnarray}
H_{\mathrm{ex-sp}} &=&\sum_{n,k_{z}}\hbar \Delta _{n,k_{z}}\widehat{a}%
_{n,k_{z}}^{\dag }\widehat{a}_{n,k_{z}}  \notag \\
&&+i\hbar \sum_{n,k_{z}}(g_{n,k_{z}}\widehat{a}_{n,k_{z}}^{\dag }\sigma
_{\downarrow \uparrow }-g_{n,k_{z}}^{\ast }\widehat{a}_{n,k_{z}}\sigma
_{\uparrow \downarrow }),
\end{eqnarray}%
where $\sigma _{ij}=\left| i\right\rangle \left\langle j\right| $($%
i,j=\uparrow ,\downarrow $) are the atomic operators; $\widehat{a}_{n,k_{z}}$
and $\widehat{a}_{n,k_{z}}^{\dag }$ are the radiation field (surface
plasmon) annihilation and creation operators;
\begin{equation}
\Delta _{n,k_{z}}=\omega _{n}(k_{z})-\omega _{0}
\end{equation}%
is the detuning of the radiation mode frequency $\omega _{n}(k_{z})$ from
the excitonic resonant \ frequency $\omega _{0}$, and $g_{n,k_{z}}=\vec{d}%
_{0}\cdot \overrightarrow{E}_{n,k_{z}}$ is the atomic field coupling. Here, $%
\vec{d}_{0}$ and $\overrightarrow{E}_{n,k_{z}}$ denote the transition dipole
moment of the exciton and the electric field, respectively. The subindex ''%
\textrm{ex-sp}'' in $H_{\mathrm{ex-sp}}$ refers to excitons (\textrm{ex})
and surface plasmons (\textrm{sp}).

Assuming that initially there is an exciton in the dot with no plasmon
excitation in the wire, the time-dependant wavefunction of the system then
has the form%
\begin{equation}
\left| \psi (t)\right\rangle =b_{e}(t)\left| \uparrow ,0\right\rangle
+\sum_{n,k_{z}}b_{n,k_{z}}(t)\left| \downarrow ,1_{n,k_{z}}\right\rangle
e^{-i\Delta _{n,k_{z}}t}.
\end{equation}%
The state vector $\left| \uparrow ,0\right\rangle $ describes an exciton in
the dot and no plasmons present, whereas $\left| \downarrow
,1_{n,k_{z}}\right\rangle $ describes the exciton recombination and a
surface plasmon emitted into mode $k_{z}$. With the time-dependent Schr\"{o}%
dinger equation, the solution of the coefficient $b_{e}(t)$ in $z$-space is
straightforwardly given by

\begin{equation}
\widetilde{b}_{e}(z)=\left\{z+\sum_{n,k_{z}}g_{n,k_{z}}g_{n,k_{z}}^{\ast }%
\frac{1}{z+i[\omega _{n}(k_{z})-\omega _{0}]}\right\}^{-1}.
\end{equation}%
In principle, $b_{e}(t)$ can be obtained by performing a numerical inverse
Laplace transformation to Eq. (6).

To grasp the main physics and without loss of generality, we focus on the
values of $\omega _{0}$ close to one particular local extremum (maximum or
minimum), where the dispersion relation for a particular mode becomes
quadratic. In this case, the dispersion relation for this particular mode
around the extreme can be approximated as
\begin{equation}
\omega _{n}(k_{z})=\omega _{n,c}\pm A_{n}(k_{z}-k_{n,c})^{2},
\end{equation}%
where the extremum is located at ($k_{n,c},\omega _{n,c}$). The $+/-$ sign
represents the approximate curve for the local minimum/maximum of the
dispersion relation. Once we make such an approximation, the radiative
dynamics of the QD exciton is just like that of a two-level atom in a
photonic crystal$^{8,15}$ with
\begin{eqnarray}
\widetilde{b}_{e}(z) &\approx &\frac{\left| \vec{d}_{0}\cdot \overrightarrow{%
E}_{n,k_{z}=k_{n,c}}\right| ^{2}}{z-\gamma /2-\frac{(-1)^{3/4}\pi }{\sqrt{%
A_{n}}\sqrt{z-i\delta }}},\text{ \ \ for local minima} \\
\widetilde{b}_{e}(z) &\approx &\frac{\left| \vec{d}_{0}\cdot \overrightarrow{%
E}_{n,k_{z}=k_{n,c}}\right| ^{2}}{z-\gamma /2+\frac{(-1)^{1/4}\pi }{\sqrt{A}%
\sqrt{z-i\delta }}},\text{ \ \ for local maxima}
\end{eqnarray}%
where
\begin{eqnarray}
\delta = \omega _{0}-\omega _{n,c}
\end{eqnarray}
is the detuning to a specific extremum and $\gamma $ is the decay rate
contributed from other modes. For example, hereafter we choose $\omega _{0}$
to be close to the minimum of the $n=1$ mode, and thus only this $n=1$, and
the $n=0$ mode, strongly interact with the exciton. The other modes can be
treated as a (Markovian) decay process with a rate $\gamma $.

The coefficient $b_{e}(t)$ can now be obtained$^{8,15}$ by performing the
Laplace transformation to Eqs.~(8,9). The black, dotted, and dashed lines in
Fig. 2(d) represent the decay dynamics of the QD excitons for different
detunings: $\delta =0.2\beta ,$ $0.4\beta ,$ $0.8\beta ,$ respectively.
Here, $\beta $ is the decay rate of the QD exciton in free space. As
mentioned above, when plotting Fig. 2(d), $\omega _{0}$ was chosen to be
close to the local minimum of the dispersion relation of the $n=1$ mode. The
radius of the wire and the wire-dot separation are identical to those in
Fig. 2(a). As can be seen in Fig. 2(d), there exists oscillatory behavior in
the decay profile of $\left| b_{e}(t)\right| ^{2}$, demonstrating that the
decay dynamics around the local extrema is non-Markovian. If one only
considers the contribution from the $n=1$ mode and set the detuning $\delta
=0$, the probability amplitude would saturate to a steady limit, as show by
the top red curve in Fig. 2(d). This \emph{quasi-dressed} state is
reminiscent of damped Rabi oscillations in cavity quantum electrodynamics,
and also appears in systems of photonic crystals$^{8,15}$.
\begin{figure}[h]
\includegraphics[width=8cm]{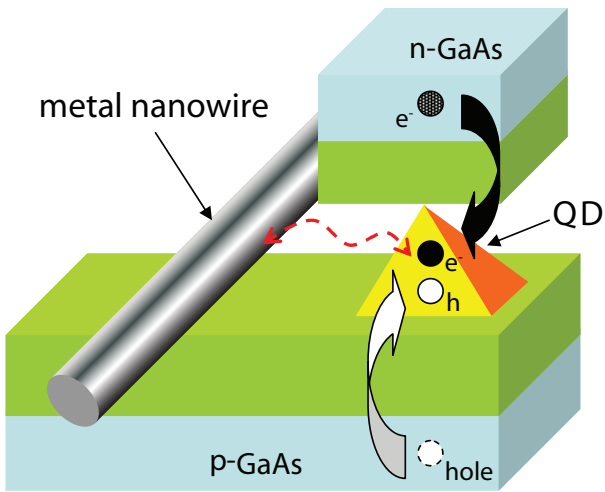}
\caption{{}(Color online) A schematic diagram of a \textrm{p-i-n} junction
with a quantum dot (QD) evanescently coupled to surface plasmons in a
nanowire. }
\end{figure}

\subsection{Readout of the band-edge effect via current-noise}

With recent advances in fabrication technologies, it is now possible to
embed QDs inside a p-i-n structure$^{6}$. Furthermore, the interest in
measurements of shot-noise in quantum transport has grown recently owing to
the possibility of extracting valuable information not available in
conventional dc transport experiments$^{16}$. We thus propose to bring these
two branches of condensed matter physics together: surface-plasmon and
current-noise measurements; i.e. by placing a QD \textrm{p-i-n} junction
close to the nanowire as shown in Fig. 3.

In addition to the Hamiltonian $H_{\mathrm{ex-sp}}$ in Eq. (4), we now need
to consider the tunnel-couplings to the electron and hole reservoirs$^{7}$:%
\begin{equation}
H_{T}=\sum_{\mathbf{q}}\left(V_{\mathbf{q}}c_{\mathbf{q}}^{\dagger }\left|
0\right\rangle \left\langle \uparrow \right| +W_{\mathbf{q}}d_{\mathbf{q}%
}^{\dagger }\left| 0\right\rangle \left\langle \downarrow \right|
+H.c.\right),
\end{equation}%
where $c_{\mathbf{q}}$ and $d_{\mathbf{q}}$ are the electron operators in
the right and left reservoirs, respectively. Here, $V_{\mathbf{q}}$ and $W_{%
\mathbf{q}}$ couple the channels $\mathbf{q}$ of the electron and the hole
reservoirs. We also introduced the three dot states: $\left| 0\right\rangle
=\left| 0,h\right\rangle $, $\left| \uparrow \right\rangle =\left|
e,h\right\rangle $, and $\left| \downarrow \right\rangle =\left|
0,0\right\rangle $, where $\left| 0,h\right\rangle $ means that there is one
hole in the QD,\ $\left| e,h\right\rangle $ is the exciton state, and $%
\left| 0,0\right\rangle $ represents the ground state with no hole and no
excited electron in the QD$^{7}$.

Together with Eq. (3), one can now write down the equation of motion for the
reduced density operator

\begin{eqnarray}
\frac{d}{dt}\rho (t) &=&-\mathrm{Tr}_{\mathrm{res}}\int_{0}^{t}dt^{\prime
}[H_{T}(t)+H_{\mathrm{ex-sp}}(t),  \notag \\
&&[H_{T}(t^{\prime })+H_{\mathrm{ex-sp}}(t^{\prime }),\widetilde{\Xi }%
(t^{\prime })]],
\end{eqnarray}%
where $\widetilde{\Xi }(t^{\prime })$ is the total density operator. Note
that the trace, $\mathrm{Tr}$, in Eq.~(12) is taken with respect to both
plasmon and electronic reservoirs. Without making the Markovian
approximation to the exciton-plasmon couplings, one can derive the equations
of motions of the dot operators$^{17}$. With the help of counting
statistics, the noise spectrum is then given by
\begin{equation}
S_{I_{R}}(\omega )=2eI\left\{1+\Gamma _{R}\left[ B(\omega )+B(-\omega )%
\right] \right\},
\end{equation}%
where
\begin{equation}
B(\omega )=\frac{A(i\omega )\Gamma _{L}}{-A(i\omega )\Gamma _{L}\Gamma
_{R}+(A(i\omega )+i\omega )(\Gamma _{L}+i\omega )(\Gamma _{R}+i\omega )}.
\end{equation}%
Here, $I$ is the stationary current, $\Gamma _{L}$ and $\Gamma _{R}$ are the
tunneling rates from the electron and hole reservoirs, and $A(z)\equiv
c(z)+c^{\ast }(z)$, where
\begin{equation}
c(z)=\sum_{n,k_{z}}\frac{g_{n,k_{z}}g_{n,k_{z}}^{\ast }}{z+i[\omega
_{n}(k_{z})-\omega _{0}]}.
\end{equation}

\begin{figure}[h]
\includegraphics[width=8cm]{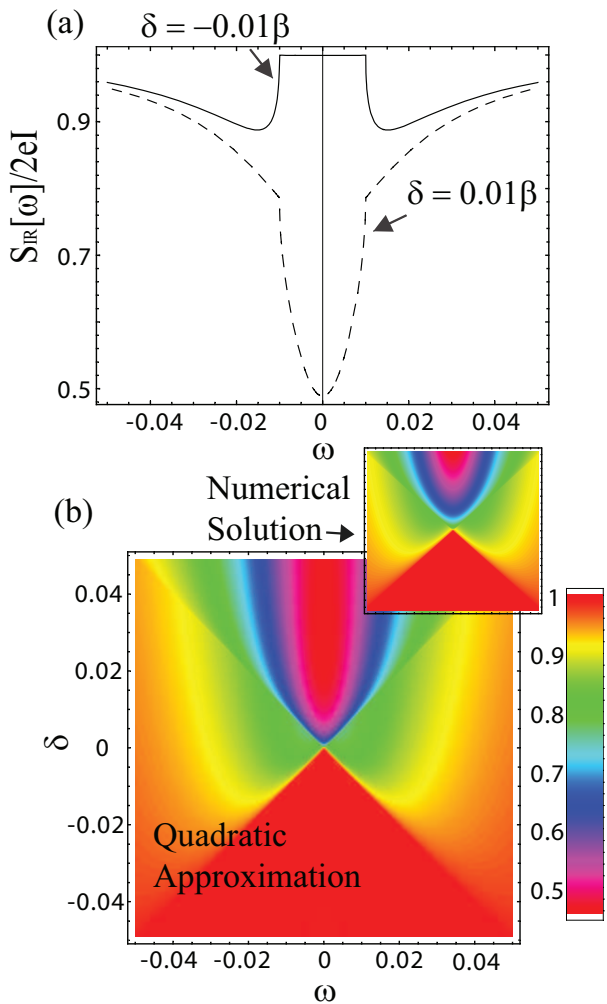}
\caption{{}(Color online) (a) Noise spectrum as a function of $\protect%
\omega $. Like that in Fig.~2(d), the value of $\protect\omega _{0}$\ here
is chosen to be close to the local minimum of the $n=1$\ mode. Here, $\Gamma
_{L}$\ and $\Gamma _{R}$\ are set equal to $0.01\protect\beta $\ and $0.1%
\protect\beta $, respectively. The solid (dashed) line represents the result
for $\protect\delta =-0.01\protect\beta $\ ($0.01\protect\beta $).\emph{\ }%
(b) Density plot of the current-noise spectrum as functions of $\protect%
\omega$ and detuning $\protect\delta =\protect\omega_0 - \protect\omega%
_{n,c} $, which are both in units of $\protect\beta $, the decay rate of the
QD in free space. As seen here, there are discontinuities along the lines $%
\protect\delta =\pm \protect\omega $, which is an indication of the
band-edge effect. In the inset we show the same calculation using the
numerical solution for the dispersion relation, not the quadratic
approximation. This illustrates that the important features are all
contained within the quadratic approximation.}
\end{figure}

Figure 4(a) shows the noise spectrum $S_{I_{R}}(\omega)$ as a function of $%
\omega $. As for Fig.~2(d), the value of $\omega _{0}$\ here is chosen to be
close to the local minimum of the $n=1$\ mode. Here, the $\Gamma _{L}$\ and $%
\Gamma _{R}$\ are set equal to $0.01\beta $\ and $0.1\beta $, respectively.
The solid (dashed) line represents the result for $\delta =\omega_0 -
\omega_{n,c}=-0.01\beta $\ ($0.01\beta $). The interesting feature here is
that there are discontinuities at $\omega =\pm 0.01\beta $. For the case of $%
\delta =-0.01\beta $, the Poissonian value of the noise spectrum [$%
S_{I_{R}}(\omega )=1$\ for $-0.01\beta <\omega <0.01\beta $] is analogous to
that of putting a two-level emitter inside the band-gap, while, for $\delta
=0.01\beta $, the sub-Poissonian value is the situation outside the band-gap$%
^{18}$. Figure 4(b) shows the density plot of the noise spectrum as
functions of both $\omega$ and detuning, $\delta =\omega _{0}-\omega _{n,c}$%
\. As seen there, for $\delta <0$, the values of $S_{I_{R}}(\omega )$\ in
the regime $-\left| \delta \right| <\omega <\left| \delta \right| $\ \ are
larger than those in $\omega <-\left| \delta \right| $\ and $\left| \delta
\right| <\omega $. For $\delta >0$, however, it is the opposite behavior. In
addition, one also observes that there are discontinuities along the lines $%
\delta =\pm \omega $. Together with the results in Fig. 4(a), we conclude
that the feature of \emph{discontinuities} in these noise spectra can
actually be viewed as an indication of a band-edge effect.

\section{Retardation effect}

By placing two QDs close to the nanowire, and by making use of the
one-dimensional propagating feature of the nanowire surface plasmons,
another non-Markovian effect, the retardation, can be observed. For
simplicity, the exciton energy $\hbar \omega _{0}$ of the two identical dots
is set well below the local minimum of the $n=1$ mode, such that only the $%
n=0$ mode contributes to the decay rate. Thus, the interaction Hamiltonian
can be expressed as

\begin{eqnarray}
\hat{H}_{I} &=&-i\hbar \sum_{l=1,2}\sum_{k_{z}}(\widehat{a}_{k_{z}}-\widehat{%
a}_{k_{z}}^{\dag })  \notag \\
&&\times (g_{k_{z}}^{\ast }e^{-ik_{z}z_{l}}\left| \downarrow \right\rangle
_{l\text{ }l}\left\langle \uparrow \right| +g_{k_{z}}e^{ik_{z}z_{l}}\left|
\uparrow \right\rangle _{l\text{ }l}\left\langle \downarrow \right| ),
\end{eqnarray}%
where $z_{l}$ is the position of the $l$-th dot, and the distance of the two
dots to the wire surface is the same. Assuming that only dot-1 is initially
excited, the state vector of the system can be written as%
\begin{equation}
\left| \psi (t)\right\rangle =b_{1}(t)\left| \uparrow \downarrow
,0\right\rangle +b_{2}(t)\left| \downarrow \uparrow ,0\right\rangle
+\sum_{k_{z}}b_{k_{z}}(t)\left| \downarrow \downarrow
,1_{n,k_{z}}\right\rangle ,
\end{equation}%
with the initial conditions: $b_{1}(0)=1$, $b_{2}(0)=0,$ and $b_{k_{z}}(0)=0$%
. The time-dependent solutions are straightforwardly given by $%
b_{1(2)}(t)=[C_{+}(t)\pm C_{-}(t)]/2$ with
\begin{equation}
C_{\pm }(t)=\frac{1}{2\pi i}\int_{-\infty +\epsilon }^{i\infty +\epsilon }ds%
\frac{e^{st}}{s+\sum_{k_{z}}\left| g_{k_{z}}\right| ^{2}[1\pm
e^{ik_{z}(z_{2}-z_{1})}]G(s)},
\end{equation}%
where
\begin{eqnarray}
G(s)=\{s+i[\omega _{n}(k_{z})-\omega _{0}]\}^{-1}+\{s+i[\omega
_{n}(k_{z})+\omega _{0}]\}^{-1}.
\end{eqnarray}
Following the well-known treatment of retardation$^{19}$, one can obtain the
probability amplitudes of the dots in the regime of $k_{0}r\geq 3$
\begin{eqnarray}
b_{1(2)}(t) &=&\!\!\!\!\!\!\!\!\sum_{%
\begin{array}{cr}
\text{{\footnotesize {m=0,2,4 ...}}} &  \\
\text{{\footnotesize {(m=1,3,5 ...)}}} &
\end{array}%
}^{\infty }\!\!\!\!\!\!\!\frac{1}{m!}(ie^{ik_{0}r})^{m}\left[\gamma
_{0}\!\left(t-\frac{mr}{v}\right)\right]^{m}  \notag \\
&&\times H\!\!\left(t-\frac{mr}{v}\right)\exp \left\{-\gamma _{0}\!\left(t-%
\frac{mr}{v}\right)\right\},
\end{eqnarray}%
where $r=\left| z_{2}-z_{1}\right| $, $k_{0}\approx \omega _{0}/v$, $v$ is
the velocity of the surface plasmon on the wire, $\gamma _{0}$ is the
spontaneous emission (SE) rate of a single QD exciton into a surface
plasmon, and $H$ is the unit step function.
\begin{figure}[h]
\includegraphics[width=8cm]{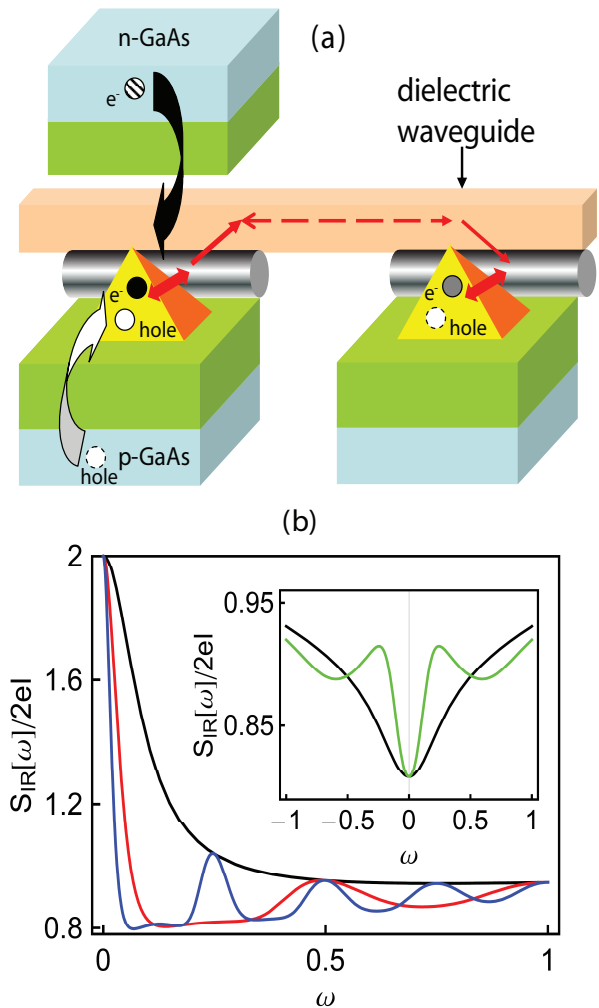}
\caption{{}(Color online) (a) Proposed device for the observation of
retardation effects via current-noise. The two QDs are coupled to two
separate nanowires. Meanwhile, the wires are evanescently coupled to a
phase-matched dielectric waveguide. (b) Current noises of the double-dot
device. The red and blue lines represent the results for $r/v=2\protect\pi $
and $4\protect\pi ,$ respectively. Recall that $r$ is the inter-dot
separation. The black line is the result for the Markovian case. The most
obvious feature for the non-Markovian effect is the oscillatory behavior
(red and blue curves). Inset: Noise spectra with (green line) and without
(black line) retardation effects when $r/v=1.9$.}
\end{figure}

One might argue that the surface plasmons inevitably experience losses as
they propagate along the nanowire, which could limit the feasibility of
observing the retardation effect. One solution to this would be to couple
two QDs to two separate nanowires. Meanwhile, the wires would be
evanescently coupled to a phase-matched dielectric waveguide$^{20}$. In this
case, one could have both the advantages of strong coupling from the surface
plasmons and also long-distance transport in the dielectric waveguide. In
addition, the non-Markovian retarding effect can also be measured via
current noise if one of the dots is embedded inside a \textrm{p-i-n}
junction as shown in Fig. 5(a). Following the procedure described above, the
noise spectrum is given by

\begin{equation}
S_{I_{R}}(\omega )=2eI\{1+\Gamma _{R}[n_{R}(s=-i\omega )+n_{R}(s=i\omega
)]\}.
\end{equation}%
In Eq.~(21), $n_{R}(s)$ is the Laplace-transformation of the ground state
occupation probability $n_{R}(t)=\left\langle \left| \downarrow \downarrow
\right\rangle \left\langle \downarrow \downarrow \right| \right\rangle _{t}$%
, where the average is over both the electronic and photonic reservoirs. The
red and blue lines in Fig.~5(b) represent the noise spectra for $\gamma
_{0}r/v=2\pi $ and $4\pi ,$ respectively. As seen there, the main difference
to the non-retarded case (black line) is the oscillatory behavior, which
depends on the inter-dot separation $r$. One recalls that in the
non-retarded situation there should be no difference whenever $\gamma
_{0}r/v=m\pi $, where $m$ is an integer$^{21}$. The green line in the inset
of Fig. 5(b) is the result for $\gamma _{0}r/v=1.9$. This means that even if
the value of $\gamma _{0}r/v$ is not equal to $m\pi $, one still could
observe the predicted oscillatory behavior.

\section{Band edge effect in phonon cavities}

The non-Markovian effects studied above can also be observed in other
physical systems. For example, if one considers a free standing slab$^{22}$
with width $w$, small elastic vibrations of a solid slab can then be defined
by a vector of relative displacement $\mathbf{u}\left( \mathbf{r},t\right) $%
. Under the isotropic elastic continuum approximation, the displacement
field $\mathbf{u}$ obeys the equation%
\begin{equation}
\frac{\partial ^{2}\mathbf{u}}{\partial t^{2}}=c_{t}^{2}\mathbf{\nabla }^{2}%
\mathbf{u}+\left( c_{l}^{2}-c_{t}^{2}\right) \mathbf{\nabla }\left( \mathbf{%
\nabla }\cdot \mathbf{u}\right) ,
\end{equation}%
where $c_{l}$ and $c_{t}$ are the velocities of longitudinal and transverse
bulk acoustic waves. To define a system of confined modes, Eq.~(22) is
complemented by the boundary conditions at the slab surface $z=\pm w/2$.
Because of the confinement, phonons will be quantized in subbands. For each
in-plane component $\mathbf{q}_{\Vert }$ of the in-plane wave vector there
are infinitely many subbands. Since two types of velocities of sound exist
in the elastic medium, there are also two transverse wavevectors $q_{l}$ and
$q_{t}$. If one further considers the deformation potential only, then there
are two main confined acoustic modes: dilatational waves and flexural waves.
For dilatational waves, the parameters $q_{l,n}$ and $q_{t,n}$ can be
determined from the Rayleigh-Lamb equation%
\begin{equation}
\frac{\tan \left( q_{t,n}w/2\right) }{\tan \left( q_{l,n}w/2\right) }=-\text{
}\frac{4q_{\Vert }q_{l,n}q_{t,n}}{(q_{\Vert }^{2}-q_{t,n}^{2})^{2}},
\end{equation}%
with the dispersion relation
\begin{equation}
\omega _{n,q_{\Vert }}=c_{l}^{2}\sqrt{q_{\Vert }^{2}+q_{l,n}^{2}}=c_{t}^{2}%
\sqrt{q_{\Vert }^{2}+q_{t,n}^{2}},
\end{equation}%
where $\omega _{n,q_{\Vert }}$ is the frequency of the dilatational wave in
mode ($n,\mathbf{q}_{\Vert }$). For the antisymmetric flexural waves, the
solutions $q_{l,n}$ and $q_{t,n}$ can also be determined by solving the
equation
\begin{equation}
\frac{\tan \left( q_{l,n}w/2\right) }{\tan \left( q_{t,n}w/2\right) }=-\text{
}\frac{4q_{\Vert }q_{l,n}q_{t,n}}{(q_{\Vert }^{2}-q_{t,n}^{2})^{2}},
\end{equation}%
together with the dispersion relation, Eq. (24).
\begin{figure}[h]
\includegraphics[width=8cm]{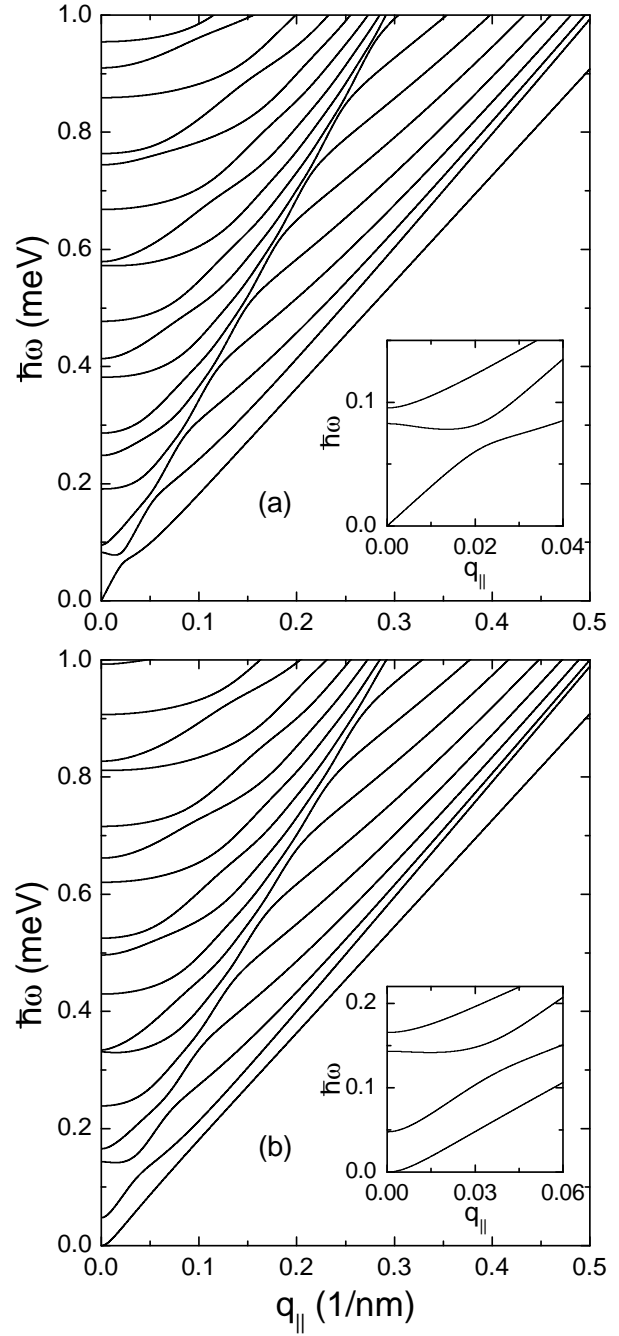}
\caption{{}Dispersion relations for (a) the dilatational waves and (b) the
flexural waves in a slab with width $w=130$ nm. The insets are the
corresponding magnified plots in the small $q_{\Vert }$ regime. As can be
seen here, the dispersion relations of the confined phonons also exhibit the
`band-edge' feature for certain values of $\protect\omega $.}
\end{figure}

Figures 6(a) and (b) numerically show the dispersion relations for
dilatational and flexural waves, respectively. As can be seen in the insets,
local minima also appear in the dispersion relations. An enhanced relaxation
rate due to the phonon van Hove singularities has been predicted if a
double-dot charge qubit$^{23}$ or single-dot spin state$^{24}$ is embedded
in such a phonon cavity. However, as we have mentioned above, the greatly
enhanced rates are also from the band-edge like effect$^{25}$, and one
should treat the dynamics of the qubits as non-Markovian. As for the
retardation effect, the two QDs may also be embedded inside a well-designed
photonic crystal waveguide$^{26}$, in which the propagation of the photon is
restricted to one dimension. In this case, the advantage of \ the
retardation effect in one dimension is still kept, and the combination with
the \textrm{p-i-n} junction should also be workable$^{18}$.

\section{Conclusions}

In summary, %based on the example of plasmons in a nanowire,
we have shown that nanowire surface plasmons, which we consider as a bosonic
reservoir with a restricted geometry, have a non-linear dispersion relation
with extreme values at certain frequencies. When coupled to a QD exciton
(combined with a \textrm{p-i-n} junction) we described how it should be
possible to observe the non-Markovian dynamics of these effects when the
recombination energy of the exciton is close to the bandgap of the plasmon
reservoir. We calculated specific results for the current-noise frequency
spectrum and observed unique signatures of these `band-edge' non-Markovian
dynamics.

Furthermore, we have shown that the retardation effect, another
non-Markovian effect which occurs when two dots are both strongly coupled to
the same nano-wire, has also unique signatures in the current-noise.
Finally, we illustrated how these effects might also be observed in a QD
spin qubit (or double-dot charge qubit) embedded inside a phonon cavity.

%We illustrated how
%the combination of \textit{p-i-n} junction with a QD exciton allows
%these non-Markovian effects to be read out via current-noise
%measurements.

\subsection{\textbf{ACKNOWLEDGMENTS}}

We would like to thank Dr.~J.~Taylor for helpful discussions. This work is
supported partially by the National Science Council, Taiwan under the grant
number 95-2112-M-006-031-MY3. FN acknowledges partial support from the
National Security Agency (NSA), Laboratory for Physical Sciences (LPS), Army
Research Office (ARO), National Science Foundation (NSF) Grant No.
EIA-0130383, JSPS-RFBR contract No. 06-02-91200, and CTC program supported
by the Japan Society for Promotion of Science (JSPS).

\end{document}